\documentclass[review]{elsarticle}

\usepackage{hyperref}
\usepackage{amssymb,amsthm}
\usepackage{amsmath}
\usepackage[mathlines]{lineno}
\usepackage{graphicx}
\usepackage{upgreek}
\usepackage{booktabs}
\usepackage[font=normalsize,labelfont=bf]{caption}
\usepackage{bm}
\usepackage{subcaption}
\usepackage{natbib} 
\usepackage{changes}
\usepackage{enumerate}
\usepackage{algorithm,algorithmic}
\biboptions{sort&compress}
\usepackage[right=1in, left=1in, top=1in, bottom=1in]{geometry} 
\hypersetup{colorlinks,linkcolor={blue},citecolor={blue},urlcolor={red}}
\usepackage{color}
\usepackage{soul}
\setstcolor{red}
\usepackage{changes}

\journal{arXiv}
\graphicspath{{./}}   

\begin{document}

\begin{frontmatter}

\title{A Method to Apply Piola-Kirchhoff Stress in Molecular Statics Simulation}

\author{Arman Ghasemi}

\author{Wei Gao\corref{mycorrespondingauthor}}
\ead{wei.gao@utsa.edu}

\cortext[mycorrespondingauthor]{Corresponding author}
\address{Department of Mechanical Engineering, University of Texas at San Antonio, San Antonio, TX 78249}

\begin{abstract}
A force-based optimization method is proposed to apply the first and second kind of Piola-Kirchhoff stresses in molecular statics simulation. This method is important for finite deformation problems in which the atomistic behavior can be more accurately described using Piola-Kirchhoff  stresses. The performance of the method is tested and validated using Silicon as a model material.   
\end{abstract}


\end{frontmatter}


\section{Introduction}

Molecular statics (MS) is widely applied in atomistic simulations under constant stress for searching local minima and the minimum energy path on the potential energy landscape. A widely used method to apply stress was proposed by Parrinello and Rahman (PR) \cite{Parrinello1981}. PR barostat was initially formulated to control stress in molecular dynamics (MD) simulations, which can be adopted to apply stress in MS simulation in the limit of zero temperature. The stress controlled by PR algorithm can be interpreted as the second Piola-Kirchhoff (PK) stress if one sets the pressure in the original PR formula to zero. Notably, the PR algorithm can be also used to apply Cauchy stress for infinitesimal deformation. For finite deformation problems, the Cauchy stress can be applied by periodically resetting the reference configuration to the current one during optimization, so that PR algorithm provides good approximation in a stepwise manner. For MD simulations, Miller et. al. \cite{Miller2016} proposed a Cauchy barostat based on a facile modification to the PR algorithm in order to apply an accurate Cauchy stress under finite deformation, where they used the target Cauchy stress as a reference to correct  the second PK stress that is controlled by PR algorithm during MD time steps.

Cauchy stress is most commonly used in MS simulations, because it measures the force per unit area in the deformed configuration and can be directly computed using Viral stress formula \mbox{\cite{Thompson2009,Admal2010}}. On the other hand, PK stresses (including the first and second kind) have been widely used in solid mechanics for finite deformation problems  \cite {Resnick2012}, which however may not be familiar to nonexperts in mechanics. Briefly, the first PK stress tensor ($\bf P$) is also called engineering stress or nominal stress, because it measures the force per unit area in reference configuration. The second PK stress tensor ($\bf S$) is defined entirely in the reference configuration: using a fictitious force pulled from the deformed configuration, which is then divided by the corresponding area in the reference configuration. One of the advantages of PK stresses is that they have well defined work conjugates, allowing accurate evaluation of the work done by a constant external stress. This has been recently exploited in nudged elastic band method for computing the barriers and minimum energy paths of solid-solid phase transitions under finite deformation \mbox{\cite{Ghasemi2019a, Ghasemi2020, Ghasemi2020a}}. Therefore, we believe that the method proposed in this paper is important to study atomistic behavior in the materials under finite or large deformation, where using PK stresses is more appropriate than Cauchy stress.

In this paper, we propose a method to apply PK stresses in MS simulation, which is different from the algorithm used in PR barostat. There are two motivations to propose a new method. First, the new method uses a force-based optimization method built upon the idea proposed by Sheppard et. al. \cite{sheppard2012}. Such method does not require an objective function, so it could be used for optimization problems when there is no a well-defined total energy, e.g. finding minimum energy path in nudged elastic band method. By contrast, PR algorithm was originally developed using extended Hamiltonian, so that the algorithm requires a well-defined energy form. In fact, the concept of the proposed algorithm has been used in our recently published finite deformation nudged elastic band method \cite{Ghasemi2019a}, while this is the first time we present the detailed implementation and discussion of the algorithm.

The second motivation of this paper is that it may not be convenient to use PR method to apply PK stresses through some publicly available software packages. This is due to the original and common implementation of PR method. The Hamiltonian in PR method (exclude the kinetic part) can be written as ${\cal H}= {\cal V} -V_{0}\left(\tilde{S}_{I J}+\tilde{p} \delta_{I J}\right) E_{I J}-\tilde{p}\left(V-V_{0}\right)$, where $\cal V$ is the potential energy, $\tilde{S}_{I J}$ is the target second PK stress tensor and $\tilde{p}$ is the target hydrostatic pressure, $E_{I J}$ is the Green strain tensor, $V_0$ and $V$ are the volumes of initial and deformed configurations, $\delta_{I J}$ represents the Kronecker delta function, and the Einstein summations is applied to the dummy index. Apparently, one can apply a hydrostatic pressure by setting $\tilde{S}_{I J}=-\tilde{p}\delta_{I J}$ to make the second term zero. In the other case, one must set $\tilde{p}=0$ in order to apply a second PK stress $\tilde{S}_{I J}$. In atomistic simulations packages (such as LAMMPS \cite{plimpton1995fast}), oftentimes, the hydrostatic and the second PK stress are not set independently, thereby the nature of the specified stress controlled by the PR algorithm is ambiguous, as discussed in \cite{tadmor2011modeling}. For example, in LAMMPS,  the stress controlled by PR algorithm is neither Cauchy stress nor the second PK stress for finite deformation.  As a result, such conventional implementation of PR algorithm makes it inconvenient to apply PK stresses, unless one does it with a customized code. To this end, one purpose of this paper is to propose an alternative and facile approach to apply PK stresses in MS simulations and to provide a ready-to-use code shared with public.

\section{Algorithm of Applying Piola-Kirchhoff  Stress}

\begin{figure}[h!]
	\centering
	\includegraphics[width=3in]{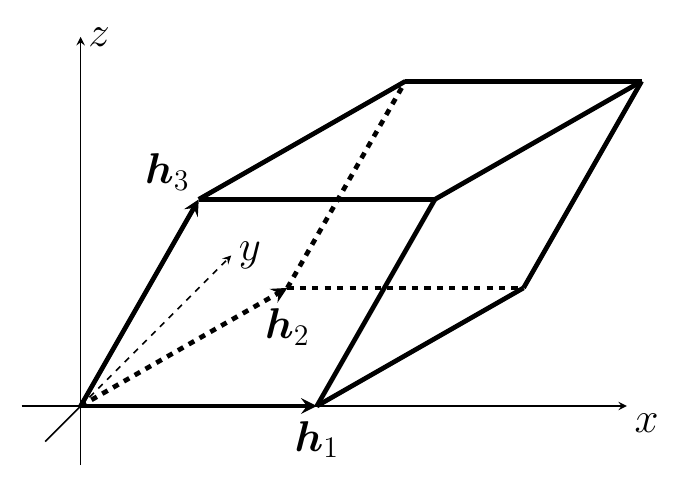}
	\caption{Computation cell and cell vectors used in the proposed algorithm.}
	\label{fig: cell}
\end{figure}

In MS simulations, the atomic degrees of freedom, i.e. atom positions, can be optimized based on the atomic forces, which converge to a given tolerance force when equilibrium is achieved. Similarly, the cell degrees of freedom, i.e. the cell vectors used to describe the deformation, can be optimized based on the stress acting on the cell. At equilibrium, the externally applied stress is balanced by the internal restoring stress computed with the Viral stress formula. To apply a stress in MS simulations, the atomic and cell degrees of freedom need to be optimized simultaneously. Sheppard et. al. \cite{sheppard2012} proposed a method to treat the atomic and cell degrees of freedom in equal footing during optimization. Following their idea, the geometry of a computation cell can be described by a cell matrix $\bf H$ written as
\begin{equation}
	{\bf H} = \left[ {\begin{array}{*{20}{c}}
			{[\bm h_1]_1}&{[\bm h_2]_1}&{[\bm h_3]_1}\\
			0&{[\bm h_2]_2}&{[\bm h_3]_2}\\
			0&0&{[\bm h_3]_3}
	\end{array}} \right],
	\label{eq: cell matrix}
\end{equation}
where $[\bm h_i]_j$ is the $j$th component of the cell vector $\bm h_i$. Particularity, $\bm h_1$ and $\bm h_2$ are confined to axis-1 and plane 1-2 as illustrated in Fig. \ref{fig: cell}. In this way, the rotational degrees of freedom of the cell are eliminated so that the cell only contains 6 independent components. The change of these components can be considered as the kinematics resulting from the corresponding Cauchy stress acting on the cell. Specifically, the six components of the Cauchy stress tensor can be expressed by three stress vectors defined as:
\begin{equation}
	\begin{aligned}	
		\bm \Sigma_1=(\sigma_{11}, 0,0),\\
		\bm \Sigma_2=(\sigma_{21}, \sigma_{22},0),\\
		\bm \Sigma_3=(\sigma_{31}, \sigma_{32},\sigma_{33}).
	\end{aligned}
	\label{eq: cauchy stress vector}
\end{equation}   
The stress vector $\bm \Sigma_i$ can be used to drive the change of the computation cell vector $\bm h_i$, along with the change of the atom positions driven by the atomic forces, leading to a deformed equilibrium configuration in terms of both atomic and cell degrees of freedom.

\begin{figure}[t]
	\centering
	\includegraphics[width=2.5in]{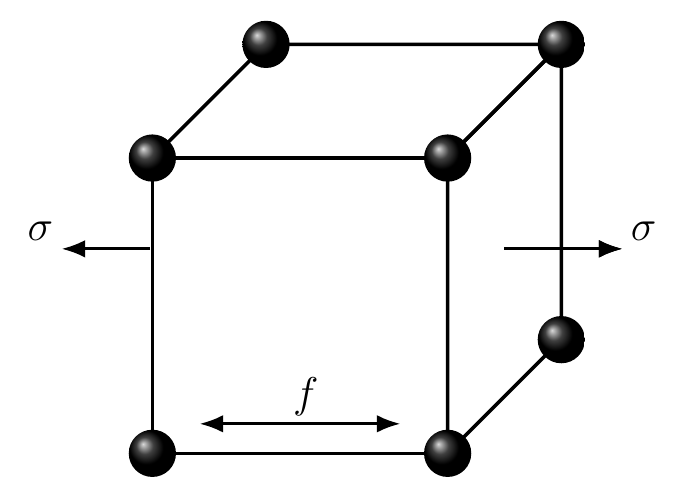}
	\caption{A simple cubic cell used to demonstrate the parameter used to scale the stress to the order of the atomic force.}
	\label{fig: cubic}
\end{figure}

Next, the algorithm is presented by considering a second PK stress $\bf S^{\text {app}}$ that is applied to a computation cell described by the matrix $\bf H$ . This stress is not directly used to drive the cell deformation but converted to a Cauchy stress during each optimization step by
\begin{equation}
	{\bm \upsigma}^{(n)} = \det({\bf F}^{(n)})^{-1} {\bf F}^{(n)} {\bf S}^{\text {app}} ({\bf F}^{(n)})^{\text T},
	\label{eq: cauchy stress tensor}
\end{equation}   
where $n$ indicates $n$th optimization step and the deformation gradient,
\begin{equation}
	{\bf F}^{(n)} = {\bf H}^{(n)} ({\bf H}^{\text {ref}})^{-1},
	\label{eq: deformation gradient}
\end{equation}
is defined with respect to a pre-defined reference cell ($\bf H^{\text {ref}}$, which could be chosen as a zero-stress configuration). Then, ${\bm \upsigma} ^{(n)}$ is used to form the stress vectors $\bm \Sigma_i^{(n)}$ defined in Eq. (\ref{eq: cauchy stress vector}). In order to update the cell vectors and atom positions simultaneously, $\bm \Sigma_i^{(n)}$ are combined with the atomic force vectors to form a generalized force vector, defined as
\begin{equation}
	\widehat{\bm f}^{(n)}=\left( \bm f_1^{(n)}, \bm f_2^{(n)}, \dots, \bm f_N^{(n)}, \alpha (\bm \Sigma_\text {1}^{\text {cell(n)}} - \bm \Sigma_\text {1}^{(n)}), \alpha (\bm \Sigma_\text {2}^{\text {cell(n)}} - {\bm \Sigma}_\text {2}^{(n)}),  \alpha (\bm \Sigma_\text {3}^{\text {cell(n)}} - {\bm \Sigma}_\text {3}^{(n)}) \right),
	\label{eq: gen force vector}
\end{equation}
where ${\bm f}_i$ is the force vector of $i$th atom of a system containing total $N$ atoms, $\bm \Sigma_i^{\text {cell}}$ is the $i$th stress vector corresponding to the internal restoring Cauchy stress ${\bm \upsigma}^{\text {cell}}$, which can be calculated by interatomic potentials or Density Functional Theory (DFT). The parameter $\alpha$ is a scaling factor to scale the stress to the order of atomic force for the convenience of convergence. A simple and intuitive choice of $\alpha$ is illustrated in Fig. \ref{fig: cubic}. Consider a simple cubic system, the stress acting on the unit cell $\sigma$ is related to the interatomic force $f$ by
\begin{equation}
	f= \left(\frac{V}{N}\right)^{\frac{2}{3}} \sigma,
\end{equation}
so the scaling factor can be taken as
\begin{equation}
	\alpha = \left(\frac{V}{N}\right)^{\frac{2}{3}}.
\end{equation}

The generalized force vector $\widehat{\bm f}$ can then be used by any force-based optimization methods to drive the change of the atomic and cell degrees of freedom. Such change is described by a generalized displacement vector
\begin{equation}
	\Delta \widehat{\bm r}^{(n)} =\left ( \Delta \bm r_1^{(n)}, \Delta \bm r_2^{(n)}, \dots, \Delta \bm r_N^{(n)}, \Delta \bm r^{*(n)}_\text {1}, \Delta \bm r_\text {2}^{*(n)}, \Delta \bm r_\text {3}^{*(n)} \right ),
	\label{eq: gen position vector}
\end{equation}
where the vector $\Delta \bm r_i$ represents the change of $i$th atom's position and $\Delta \bm r_i^*$ represents the generalized displacement  that is used to update the cell vectors by
\begin{equation}
	\Delta \bm h_i = \beta \Delta \bm r_i^*.
	\label{eq: cell vector change}
\end{equation} 
A scaling factor, $\beta$, is introduced to scale $\Delta \bm r_i^*$ to the order of cell vector $\Delta \bm h_i$. $\beta$ can be chosen as
\begin{equation}
	\beta = \left ( N \right) ^{\frac{1}{3}},
\end{equation}   
if we consider the simple cubic system shown in Fig. \ref{fig: cubic}. Once the cell vectors are changed, the deformation gradient ${\bf F}^\text{(n)}$ and the Cauchy stress ${\bm \upsigma}^{(n)}$ can be respectively updated by Eq. (\ref{eq: deformation gradient}) and Eq. (\ref{eq: cauchy stress tensor}). Subsequently, the updated generalized force vector $\widehat{\bm f}^{(n)}$ drives another change on atom positions and cell vectors, generating an iterative process, which is converged until the force vectors of all  elements inside $\widehat{\bm f}^{(n)}$ are less than a given tolerance $f_\text {max}$, namely
\begin{equation}
	\max_{i} \left\Vert {\widehat{\bm f}^{(n)}_i}\right \Vert < f_\text{max}. 
\end{equation}
Based on our numerical tests, a slight modification to the values of scaling factors $\alpha$ and $\beta$ (for example, multiplying them by 2) is not likely to jeopardize the convergence of the algorithm, however, it may lead to a different convergence rate. Following the same algorithm described above, a first PK stress ${\bf P}^{\text {app}}$ can be also applied by replacing Eq. (\ref{eq: cauchy stress tensor}) with
\begin{equation}
	{\bm \upsigma}^{(n)} = \det({\bf F}^{(n)})^{-1}  {\bf P}^{\text {app}} ({\bf F}^{(n)})^{\text T}.
	\label{eq: cauchy stress tensor by 1st PK}
\end{equation}   
 
 \begin{algorithm}[b!]
 	\caption{Apply Piola-Kirchhoff stress using MDmin optimizer}
 	\label{alg}		
 	\begin{algorithmic}[1]
 		\STATE initialize $\widehat{\bm f}^{(0)}$ defined in Eq. (\ref{eq: cauchy stress tensor}) 	
 		\STATE  {set the initial velocity $\widehat{\bm v}^{(0)}$ defined in Eq. (\ref{eq: gen velocity vector}) to zero}
 		\STATE{$n=0$}
 		\WHILE{$\max_{i} \left\Vert {\widehat{\bm f}^{(n)}_i}\right \Vert \geq f_\text{max} $}
 		\STATE {compute deformation gradient ${\bf F}^{(n)}$ with Eq. (\ref{eq: deformation gradient})}
 		\STATE{compute applied Cauchy stress ${\bm \upsigma}^{(n)}$ with Eq. (\ref{eq: cauchy stress tensor})} 
 		\STATE {compute atomic forces ${\bm f}^{(n)}_i $ and Cauchy stress ${\bm \upsigma}^{\text {cell}}$ from emipirical potentials or DFT}
 		\STATE{form generalized force vector $\widehat {\bm f}^{(n)}$ using Eq. (\ref{eq: gen force vector})}
 		\STATE{$\widehat {\bm v}^{(n+\frac{1}{2})} = \widehat {\bm v}^{(n)}+ \widehat {\bm f}^{(n)} \Delta t/2$ \hspace{2mm}}
 		\IF{$\widehat {\bm f}^{(n)} \bm{\cdot} \widehat {\bm v}^{(n+\frac{1}{2})} < 0 $}
 		\STATE $\widehat {\bm v}^{(n+\frac{1}{2})} = 0$
 		\ELSE
 		\STATE{$\widehat {\bm v}^{(n+\frac{1}{2})} = \widehat{\bm f}^{(n)} \dfrac{\widehat{\bm f}^{(n)} \bm {\cdot} \widehat {\bm v}^{(n+\frac{1}{2})}} {\widehat{\bm f}^{(n)} \bm{\cdot} \widehat{\bm f}^{(n)} }$}
 		\ENDIF
 		\STATE {$\widehat {\bm v}^{(n+1)} = \widehat {\bm v}^{(n+\frac{1}{2})}+ \widehat {\bm f}^{(n)} \Delta t/2$}
 		\STATE compute $\Delta \widehat{\bm r}^{(n)}$ defined in Eq. (\ref{eq: gen position vector}): $\Delta \widehat{\bm r}^{(n)} = \widehat {\bm v}^{(n+1)} \Delta t$
 		\STATE update atomic positions ${\bm r}^{(n+1)}_i = {\bm r}^{(n)}_i + \Delta {\bm r}^{(n)} $ and cell vectors ${\bm h}^{(n+1)}_{k} = {\bm h}^{(n)}_{k} + \beta \Delta {\bm r}^{*(n)}$
 		\STATE{$n = n+1$}
 		\ENDWHILE
 	\end{algorithmic}
 \end{algorithm} 

The algorithm described above can be integrated to many optimization methods, such as steepest descents and conjugate gradient methods ((both have to be modified to work with forces) as well as damped dynamics.  Here, we use the MDmin optimization method, as implemented in Atomic Simulation Environment (ASE) package \cite{HjorthLarsen2017a}, to demonstrate a detailed implementation of the algorithm.  MDmin is a damped dynamics routine where the damping parameter is replaced by a projection of the velocity along the force direction. It is simply a modification of the Velocity Verlet molecular dynamics algorithm. In addition, the conventional velocity is generalized in order to include the cell degrees of freedom,
\begin{equation}
\widehat{\bm v}^{(n)}=\left (\bm v_1^{(n)}, \bm v_2^{(n)}, \dots, \bm v_N^{(n)}, \bm v_{1}^{*(n)}, \bm v_{2}^{*(n)}, \bm v_{3}^{*(n)} \right ),
\label{eq: gen velocity vector}
\end{equation}   
where $\bm v_i$ is the velocity of $i$th atom and $\bm v^*_i$ represents the generalized velocity induced by the generalized forces. The procedure of applying a second PK stress is summarized in Algorithm \ref{alg}, where the MDmin method is applied from line 9 to line 16.  At each time step, the dot product between the forces and the velocity vectors is checked. If it is zero, the velocity is set to zero, otherwise, the velocity is projected to the force direction and its magnitude is set equal to the damping parameter. The atomic and cell degrees of freedom are both updated by Velocity Verlet.  The MDmin method can perform very efficiently for large systems because it takes advantage of the physics of the problem.

\begin{figure}[b!]
	\centering
	\begin{subfigure}[b]{0.4\textwidth}
		\centering
		\includegraphics[width=\textwidth]{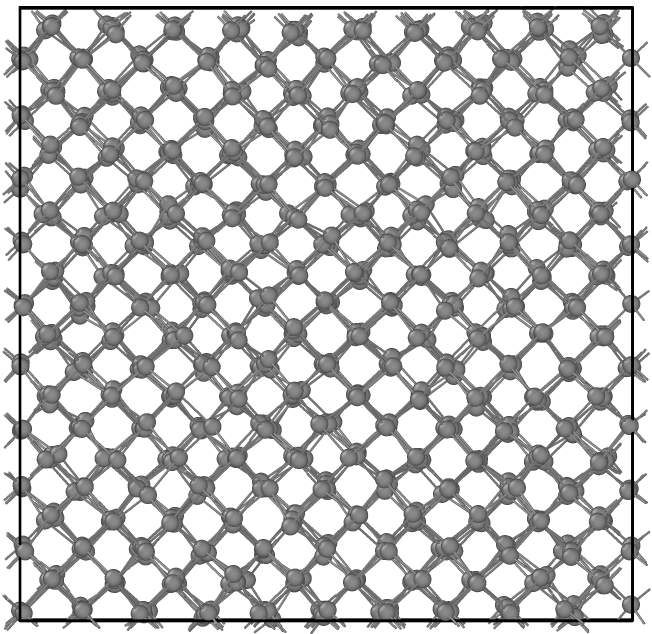}
		\caption{}
		\label{fig: silicon-a}
	\end{subfigure}
	\hfill
	\begin{subfigure}[b]{0.4\textwidth}
		\centering
		\includegraphics[width=\textwidth]{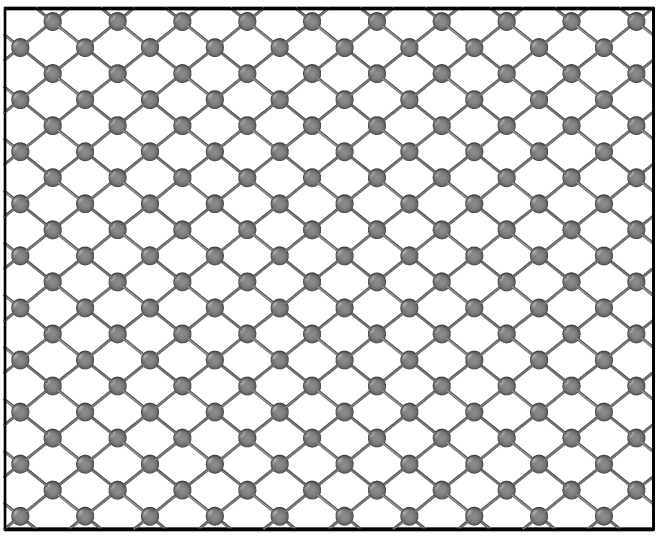}
		\caption{ }
		\label{fig: silicon-b}
	\end{subfigure}
	\caption{The computation cell of diamond cubic Si containing 1000 atoms. (a) Initially, all atoms are randomly perturbed from their equilibrium positions. (b) When PK stress is applied on [001] direction as in stress state (\ref{case 1}), both atoms and cell are optimized to the equilibrium configuration.}
	\label{fig: silicon}
\end{figure}

\section{Numerical Example}
In this section, we use diamond cubic silicon as a model material to demonstrate the performance and application of the proposed algorithm. Fig. \ref{fig: silicon-a} shows a computation cell containing 1000 atoms which are randomly disturbed from their equilibrium positions. In this way, both the atoms and the cell are initially set to non-equilibrium state. A zero stress equilibrium configuration is taken as the reference configuration for measuring PK stresses. Two stress states are applied in this example (units: GPa, unspecified stress components are zeros): 
\begin{enumerate}[(i)]
	\item \label{case 1} uniaxial compression: the first PK stress $P_{33} = -13.452$; the second PK stress $S_{33} = -15.823$, which are both equivalent to a Cauchy stress $\sigma_{33} = -12$.
	\item \label{case 2} compression plus shear: the first PK stress $P_{11}=-0.620$, $P_{12} = 5.653$, $P_{21} = 5.703$ and $P_{33} = -7.401$; the second PK stress $S_{11}=-1.211$, $S_{12} = S_{21} = 5.522$ and $S_{33} = -8.042$, which are both equivalent to Cauchy stress $\sigma_{12} = \sigma_{21} = 6$ and $\sigma_{33} = -7$. 
\end{enumerate}

\begin{figure}[h!]
	\centering
	\includegraphics[width=6in]{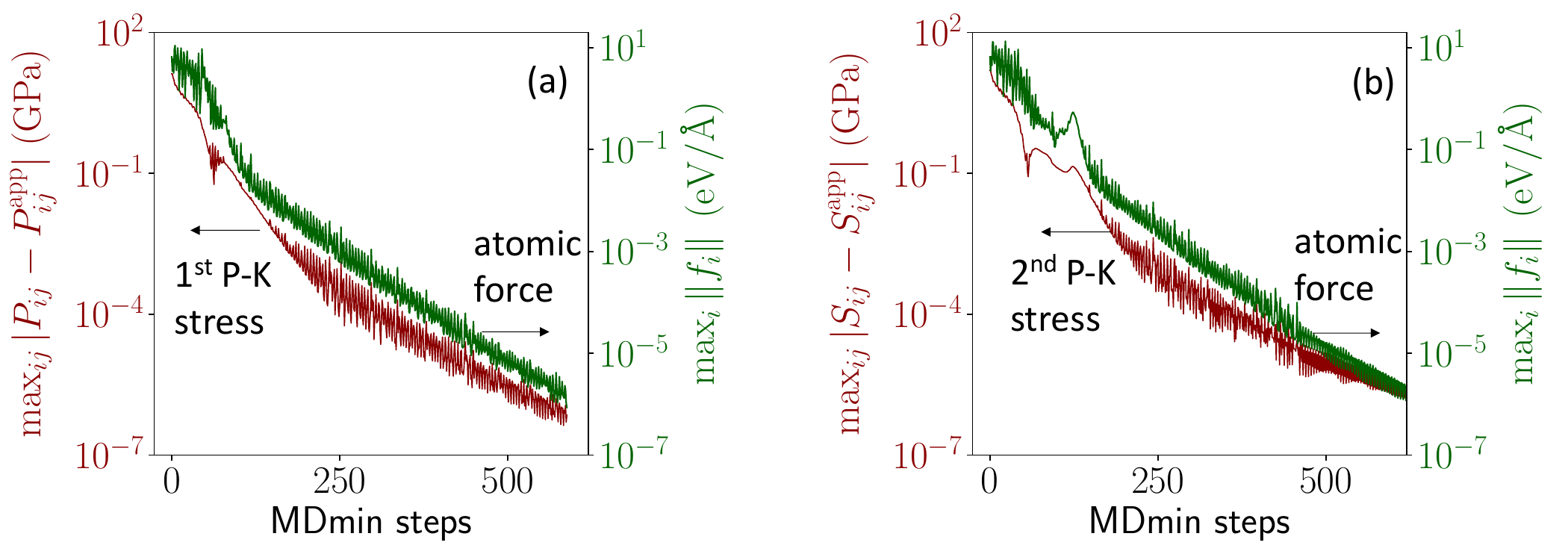}
	\caption{Convergence of the atomic forces and stress under uniaxial compressive stress specified by (\ref{case 1}) in the text, (a) for first PK stress and (b) for second PK stress.}
	\label{fig: convergence compression}
\end{figure}

\begin{figure}[h!]
	\centering
	\includegraphics[width=6in]{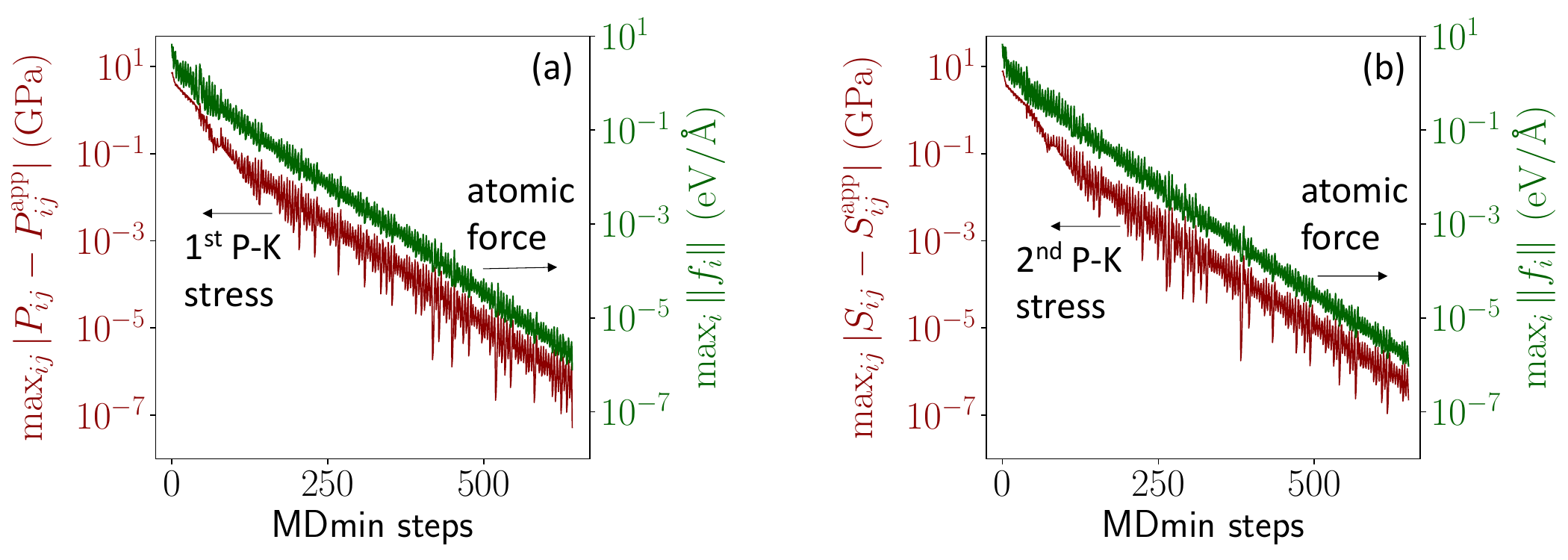}
	\caption{Convergence of the atomic forces and stress under compressive and shear stresses specified by (\ref{case 2}) in the text, (a) for first PK stress and (b) for second PK stress.}
	\label{fig: convergence shearpress}
\end{figure}

All calculations are performed with Stillinger-Weber (SW) \cite{Stillinger1985a} interatomic potential as implemented in LAMMPS \cite{plimpton1995fast}. The convergence of both the atomic forces and stresses are monitored during the optimizations, as shown in Fig. \ref{fig: convergence compression} and Fig. \ref{fig: convergence shearpress} respectively for stress states (\ref{case 1}) and (\ref{case 2}). It can be seen that the atomic forces and stresses converge at very similar rates, meaning that the atomic and cell degrees of freedom are treated equivalently during optimization. In addition, a similar convergence behavior is shown for the first and second PK stress, because they are both converted to the Cauchy stress before being passed to the optimizer. It is also confirmed that the optimizations under the first and second PK stress prescribed in (\ref{case 1}) and (\ref{case 2}) yield the correct configurations where both the atoms and the computation cell are brought to the equilibrium states, as shown in Fig. \ref{fig: silicon-b}.         

The numerical examples and the proposed algorithm are implemented based on the Atomic Simulation Environment (ASE) \cite{HjorthLarsen2017a}, an open source Python package. The advantage of ASE is that it provides an interface to various external atomistic computational codes, such as VASP and LAMMPS, which can be used as the calculators to compute atomic forces and stresses. The code and the example scripts reported in this paper are available at: \url{https://github.com/Gao-Group/stressbox}.

\begin{figure}[h!]
	\centering
	\includegraphics[width=3in]{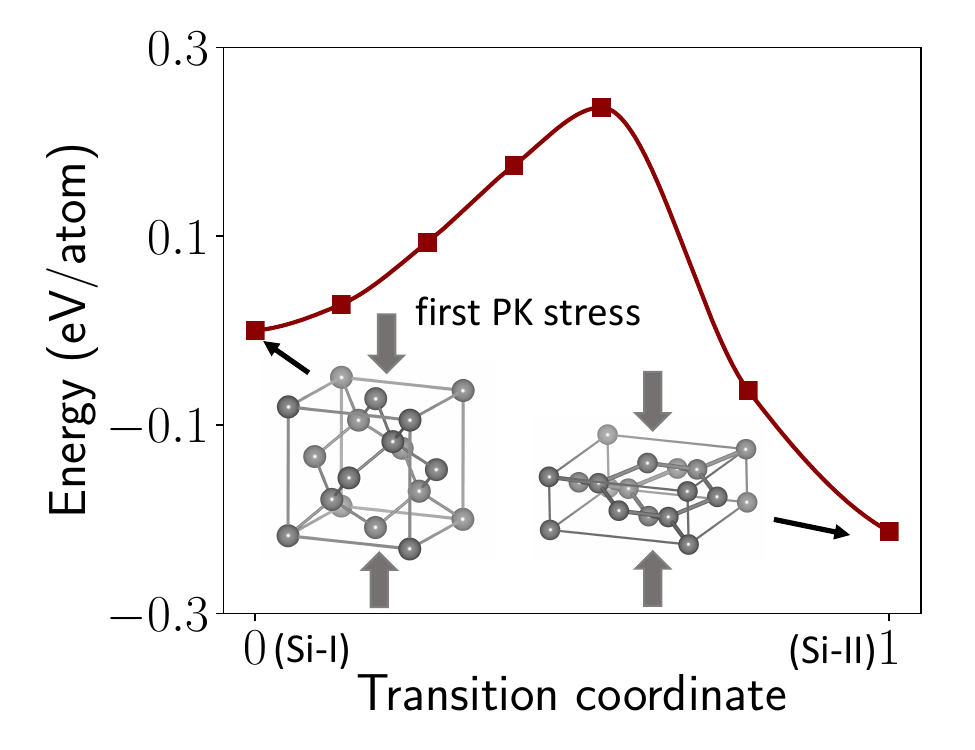}
	\caption{The minimum energy path of Si-I to Si-II phase transition under 8 GPa compressive first PK stress applied along [001] direction.}
	\label{fig: mep}
\end{figure}

Finally, we use the phase transition of Silicon, from a diamond cubic (Si-I) phase to a metallic $\beta$-tin structure (Si-II) \cite{wippermann2016novel}, as an example to explain the importance of applying PK stresses in MS simulations. The transition from Si-I to Si-II is accompanied with finite lattice deformation, and the work done by the external stress contribute significantly to the transition barriers and the minimum energy path (MEP). As mentioned in the introduction, using Cauchy stress yields inaccurate evaluation of the work done by the stress, and hence lead to inaccurate barriers and deviated MEP. Because of this, PK stresses are better suited for phase transition problems when material is subjected to finite deformation. Recently, we proposed a finite deformation nudged elastic band (FD-NEB) method to compute the transition barrier and MEP \cite{Ghasemi2019a} under a constant PK stress. In order to compute the barrier and MEP of Si-I to Si-II phase transition, one important step is to apply the PK stress to both the initial state Si-I and the final state Si-II using the algorithm described above. After that, a number of intermediate states generated between the initial and final states are optimized simultaneously under the applied PK stress until a converged MEP is established. A typical MEP calculated under 8 GPa compressive first PK stress is shown in Fig. \ref{fig: mep}. SW interatomic potential is used in this calculation, which overestimates the phase transition barriers comparing to DFT results, as noted by previous studies \cite{Zarkevich2018}.   

\section {Summary}
A new method is formulated to apply the first and second kind of PK stresses in MS simulation. The proposed force-based algorithm can be integrated to a variety of optimization methods. A damped dynamics optimizer, MDmin, is used to demonstrate the implementation of the proposed algorithm. The performance of the method is tested on diamond cubic silicon material, showing that the atomic and cell degrees of freedom can be optimized equivalently under constant PK stresses. The method is useful for finite deformation problems in which PK stresses are more appropriate to describe the atomic behavior, such as the phase transitions in the materials subjected to finite deformation.


\section*{Acknowledgments}
The authors gratefully acknowledge financial support of this work by the National Science Foundation through Grant no.
CMMI-1930783. This project was funded (in-part) by the University of Texas at San Antonio, Office of the Vice President for Research, Economic Development \& Knowledge Enterprise. The authors acknowledge the Texas Advanced Computing Center (TACC) at the University of Texas at Austin for providing HPC resources that have contributed to the research results reported within this paper.

\bibliographystyle{elsarticle-num}
\bibliography{stressbox}

\end{document}